\documentclass[%
 aip,
 amsmath,amssymb,
 reprint,%
]{revtex4-1}

\usepackage{graphicx}
\usepackage{dcolumn}
\usepackage{bm}
\usepackage[utf8]{inputenc}
\usepackage[T1]{fontenc}
\usepackage{mathptmx}
\usepackage{etoolbox}

\usepackage{xcolor}
\newcommand{\rv}[1]{\textcolor{black}{#1}}

\makeatletter
\def\@email#1#2{%
 \endgroup
 \patchcmd{\titleblock@produce}
  {\frontmatter@RRAPformat}
  {\frontmatter@RRAPformat{\produce@RRAP{*#1\href{mailto:#2}{#2}}}\frontmatter@RRAPformat}
  {}{}
}%
\makeatother
\begin{document}

\preprint{AIP/123-QED}

\title[]{Integrating planar circuits with superconducting 3D microwave cavities using tunable low-loss couplers}

\author{Ziyi Zhao} \thanks{Author to whom correspondence should be addressed: zizh9113@colorado.edu}
    \affiliation{JILA, National Institute of Standards and Technology and the University of Colorado, Boulder, CO, USA.}
    \affiliation{Department of Physics, University of Colorado, Boulder, CO, USA.}
\author{Eva Gurra}
    \affiliation{JILA, National Institute of Standards and Technology and the University of Colorado, Boulder, CO, USA.}
    \affiliation{Department of Physics, University of Colorado, Boulder, CO, USA.}
\author{Eric I. Rosenthal}
    \affiliation{E. L. Ginzton Laboratory, Stanford University, Stanford, California 94305, USA}
\author{Leila R. Vale}
    \affiliation{National Institute of Standards and Technology, Boulder, Colorado 80305, USA}
\author{Gene C. Hilton}
    \affiliation{National Institute of Standards and Technology, Boulder, Colorado 80305, USA}
\author{K. W. Lehnert}
    \affiliation{JILA, National Institute of Standards and Technology and the University of Colorado, Boulder, CO, USA.}
    \affiliation{Department of Physics, University of Colorado, Boulder, CO, USA.}
    \affiliation{National Institute of Standards and Technology, Boulder, Colorado 80305, USA}

\date{\today}

\begin{abstract}

We design and test a low-loss interface between superconducting 3-dimensional microwave cavities and 2-dimensional circuits, where the coupling rate is highly tunable.
This interface seamlessly integrates a loop antenna and a Josephson junction-based coupling element.
We demonstrate that the loss added by connecting this interface to the cavity is \rv{1.28 kHz, corresponding to an inverse quality factor of $1/(4.5 \times 10^6)$}.
Furthermore, we show that the cavity's external coupling rate to a 50 $\Omega$ transmission line can be tuned from negligibly small to over 3 orders of magnitude larger than its internal loss rate in a characteristic time of 3.2 ns.
This switching speed does not impose additional limits on the coupling rate because it is much faster than the coupling rate.
Moreover, the coupler can be controlled by low frequency signals to avoid interference with microwave signals near the cavity or qubit frequencies.
Finally, the coupling element introduces a 0.04 Hz/photon self-Kerr nonlinearity to the cavity, remaining linear in high photon number operations.

\end{abstract}

\maketitle

To scale up and perform more demanding quantum information processing tasks, modular quantum networks built upon superconducting circuits have emerged as a promising approach \cite{multilayer}.
This architecture distributes entanglement over spatially separated modules to divide the scaling challenge into independent and more manageable parts \cite{Chou_2018, Zhou_2021, Magnard_2020, niu_low-loss_2023,kannan_waveguide_2023}. 
In order for these modules to store quantum states in high quality factor 3-dimensional microwave cavities, an ongoing challenge is to implement rapid, reconfigurable, and low-loss swap interactions among the cavities \cite{Wang_2016, Reagor_2016, Chakram_2021}. 
Because it is hard to apply low frequency control signals to the coupling elements into 3D cavities without spoiling their quality factor, one solution uses microwave-actuated nonlinear elements and coaxial cable modes to couple 3D cavities \cite{Pfaff_2017, Luke_2021, Axline_2017, Peter_2022}.
In contrast, by routing the microwave signal out of the superconducting enclosure and onto a planar circuit, where the mode is confined between the circuit elements on the chip and the flat ground plane beneath them, it is possible to use a much wider class of circuit elements, including those with multiple wiring layers, magnetic flux bias, and low frequency control.
Furthermore, the characteristics of the planar circuit elements in such devices can be accurately simulated using the well developed planar method of moment solvers.
It is then desirable to incorporate the design flexibility and large scale manufacturability of the planar circuits with high quality factor 3-dimensional microwave cavities with an integrated interface.

Here, we demonstrate such an interface that seamlessly integrates a loop antenna and a tunable coupling element on a planar circuit and connect it to a coaxial quarter wave cavity. 
In our design, we utilize a symmetric coupling element connected to a superconducting loop, which is an antenna coupled to the magnetic field of the cavity \cite{TIB1}.
The integrated nature of this interface allows us to precisely predict all the microwave modes in the system and eliminate seam loss in the microwave signal lines \cite{Demonstration_Brecht}.
To this end, our device can tune the external coupling rate of the cavity by a factor of 1130 compared to its internal linewidth, while introducing a loss of \rv{1.28 kHz}. This operation can be performed with a characteristic time of $3.2$ ns and remains linear at high intracavity photon number \cite{Campagne-Ibarcq2020}.


\begin{figure*}
	\centering
	\includegraphics[width = 510pt]{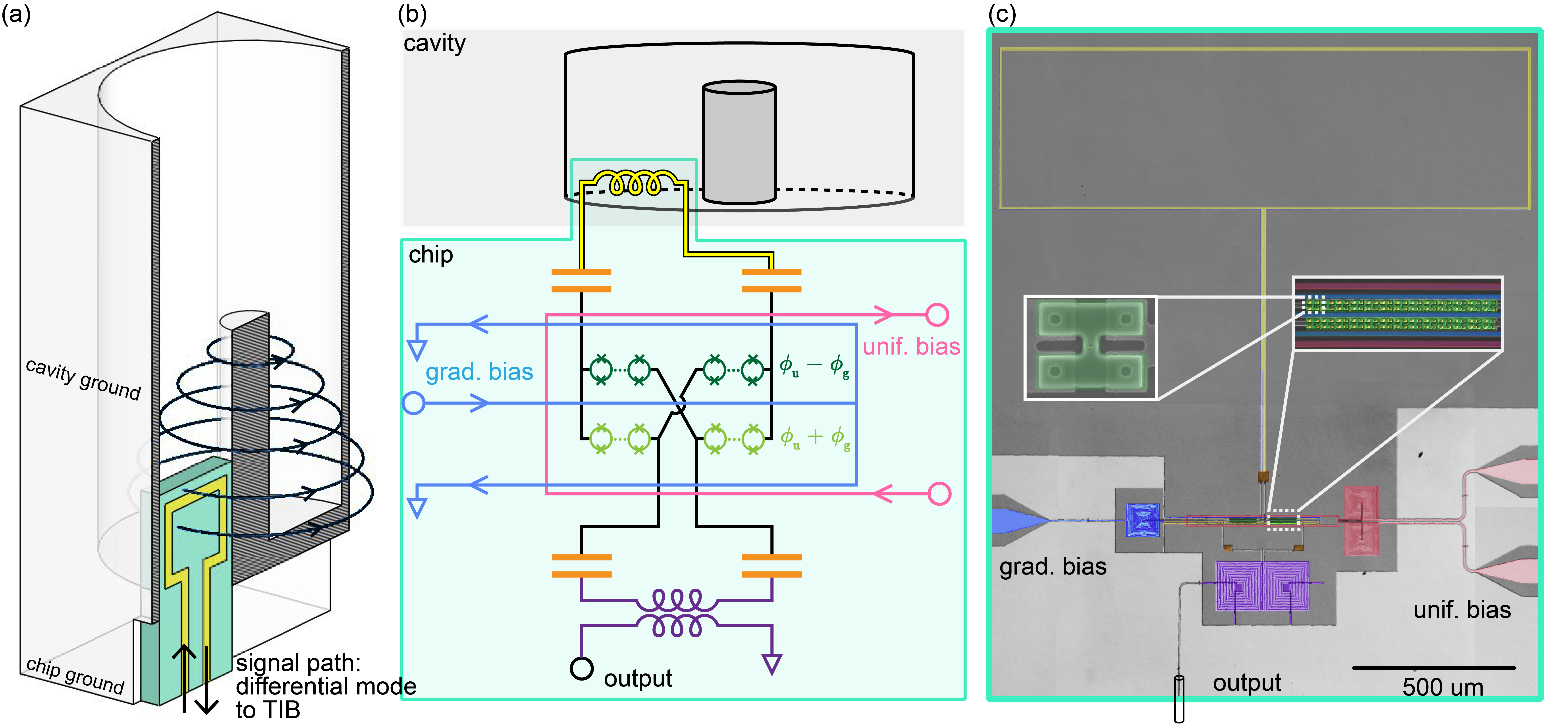}
	\caption{Device description and image
	a) A superconducting loop ($3.5\ $mm$ \times 1.0\ $mm) (yellow) on a Si substrate (turquoise) is inductively coupled to a 3D microwave cavity made of high purity (5N) aluminum ($\omega_{\text{cav}}/2\pi$ = 5.772 GHz) (white), which also acts as the ground plane for the microstrip lines on the chip \rv{(see supplementary material)}. 
	The magnetic field is depicted as the black contour lines with arrows surrounding the post, where their distance to the post signifies their amplitude. 
	b) An electrical schematic of the system shows that the cavity is connected to the coupling element through a superconducting loop (yellow) and capacitors (orange).
	The coupling element is a Wheatstone bridge, consisting of SQUIDs arrays.
	The dark (light) green SQUID arrays experience the same flux bias and have the same inductance, but the flux bias and inductance are different between dark and light green SQUID arrays.
	The uniform (pink) and gradiometric (blue) bias line are used to control the sum and difference flux bias.
	The output is converted to single-ended mode by a balun (purple). 
	c) A false-color image of the chip, fabricated with the NIST niobium trilayer process \cite{Demonstration_Mates}, shows the superconducting loop and the coplanar strip (yellow) connecting it to the inductive Wheatstone bridge,
	gradiometric bias line with filter (blue), 
	uniform bias line with filter (pink), 
	balun (purple), 
	SQUID arrays (green), 
	capacitors (orange), 
    and on-chip ground (light grey).
    An on-chip ground is only used for the control lines or for the signal path after it has left the interface via the output balun.
    The insets show an array of SQUIDs, and a zoomed in picture of 2 SQUIDs.
 }
	\label{fig:1}
\end{figure*}

The symmetry of the coupling element, a Tunable Inductor Bridge (TIB), provides a low residual coupling in its off-mode (balanced) configuration, and consequently, the cavity off-mode linewidth is dominated by its internal loss rate \cite{TIB1, Eric_2021,Kerckhoff_2015}. 
This symmetry is preserved throughout the signal path, which is composed of the TIB, a coplanar strip transmission line, and the loop antenna, a superconducting loop that is inserted into the shorted end of a quarter-wave coaxial resonator (Fig.~\ref{fig:1}).
The oscillating magnetic field of the cavity's resonant mode launches a differential mode to the coplanar strip transmission line (Fig.~\ref{fig:1}a).

In addition to high on/off ratio, the integrated nature of this interface makes it low loss; it eliminates places where the current flows between disjoint conductors (seams).
The propagating mode passing through the interface is a conductor-backed coplanar strip mode, where the signal flows on the metal traces on the chip, and the conductor backing is machined from the same piece of aluminum as the cavity. In the signal path, the loop antenna and the coupling element are already on the same chip and require no on-chip ground reference (Fig.~\ref{fig:1}c).
Without extra microwave components, for instance, microwave connectors, cables, printed circuit boards or wirebonds to connect the antenna and the coupling elements, the electrical length between them is minimized and the parasitic microwave modes can be predicted during the design process.

The tunability of this interface is achieved by controlling the TIB, which is an inductive Wheatstone bridge \cite{Bergeal_JRM_2010} using Superconducting Quantum Interference Device (SQUID) arrays that act as flux tunable inductors whose non-linearity is suppressed by increasing the number of SQUIDs in the array (Fig.~\ref{fig:1}b). 
The on/off mode of the coupler is controlled by the currents in two bias lines, one gradiometric and one uniform \rv{(see supplementary material)}, and the net flux they induce in the SQUID loop is shown in Fig.~\ref{fig:1}b.
This arrangement causes two inductors on opposite sides of the bridge to experience the same flux and therefore have the same inductance, but neighboring inductors have different inductance, as indicated by the different shading of the SQUID symbols.
At zero gradiometric bias, all inductances are equal, the bridge is balanced, and the coupler is off.
Because the uniform bias only needs to be set once in the experiment, in the following characterization, only the gradiometric bias $\Phi$ is tuned.
The external coupling rate can be tuned continuously, and this can be used in pulse shaping schemes, but here we only characterize the ``digital'' behavior of our device, where the coupling element is switched between the on and off states.
Because the cavity resonance frequency tunes with the bias, operating digitally allows us to work with a simplified picture where the spectrum of the outgoing microwave field from the cavity is always centered at the cavity frequency when the switch is on.

\begin{figure}
	\includegraphics[width = 246 pt]{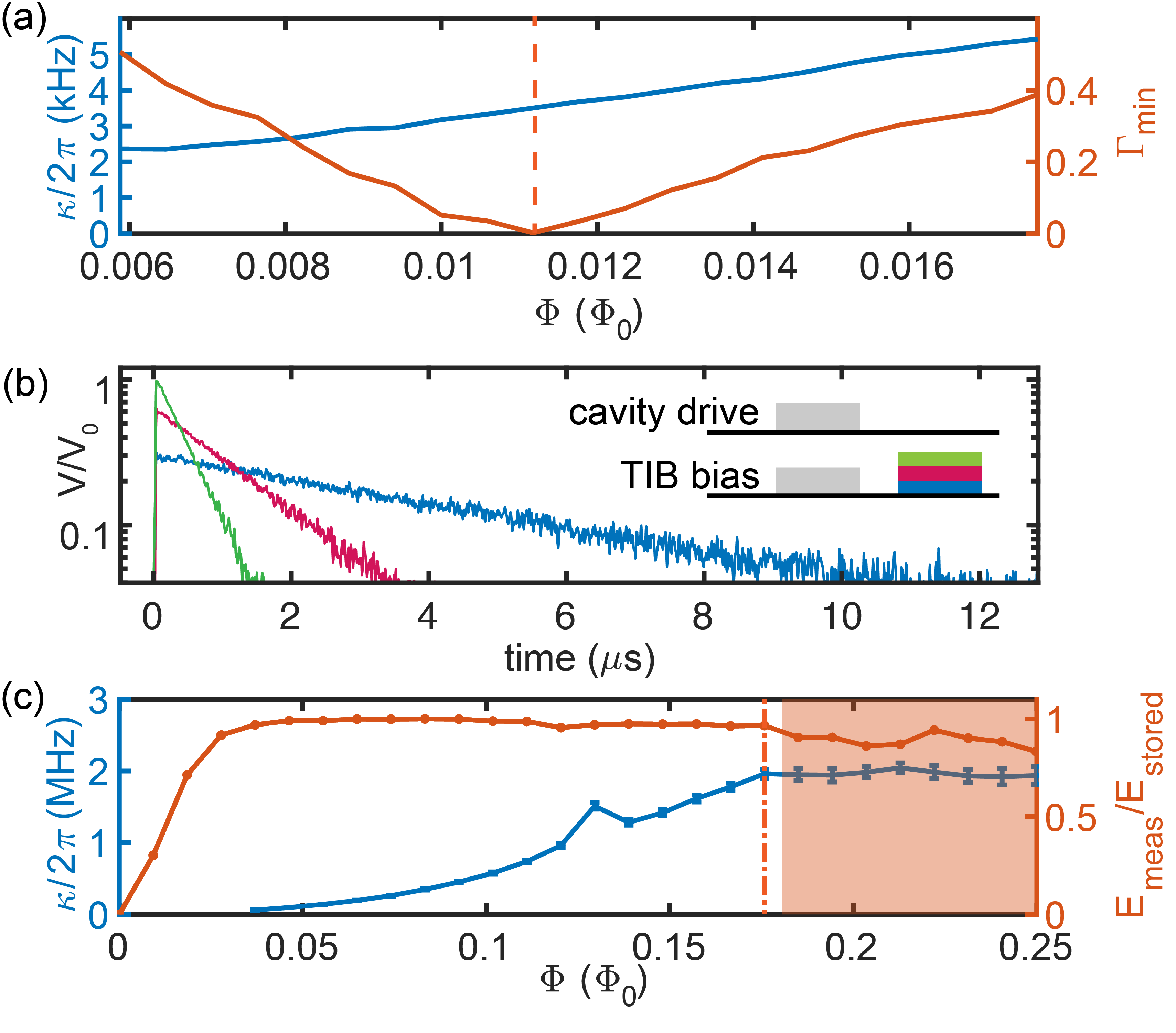}
	\caption{Coupling rate measurements
	a) Total linewidth $\kappa$ (blue) and minimal reflection amplitude $|\Gamma_{\text{min}}|$ (solid orange) are extracted as a function of bias near the critical coupling point (dashed orange).
	b) Representative samples of the time traces are shown with different decay rates. Each trace is normalized by the maximum of the green trace $V_0 = \sqrt{E_{\text{stored}}~\kappa_g~Z_0}$, where $\kappa_g$ is the total linewidth at the green trace bias and $Z_0$ is the characteristic impedance of the transmission line connected to the output port.
    The inset is the time sequence of excitations applied to the cavity and control signals to the TIB. The different bias values correspond to the time traces with the same colors.
	c) The total decay rate (blue) and measured energies ($E_{\text{meas}} \propto \int dt V^2$, solid orange) are extracted from time domain measurements. 
    $E_{\text{stored}}$ is the maximum of $E_{\text{meas}}$ over all bias points.
    The bias with maximal achievable coupling rate is shown as dotted dashed orange.
    The shading indicates the region of larger bias where energy appears to be lost to other modes.}
	\label{fig:2}
\end{figure}

When the coupler is off, the high quality factor of the cavity should be preserved by minimizing the loss introduced by the chip.
In Fig.~\ref{fig:2}a, we determine this loss by measuring the reflection coefficient $\Gamma$ from the output port, using a signal that puts approximately 1000 photon into the cavity. We sweep the bias close to critical coupling, where the reflection amplitude is minimal and the corresponding linewidth is exactly twice the internal loss rate.
We observe that the total cavity linewidth is $\kappa/2\pi = $ 3.46 $\pm$ 0.02 kHz, indicating the internal loss rate is $\kappa_{\text{int}}/2\pi = $ 1.73 $\pm$ 0.01 kHz. 
Subtracting the 0.45 $\pm$ 0.01 kHz loss rate of the cavity itself, measured separately without inserting the chip, the introduced loss rate is about 1.28 $\pm$ 0.01 kHz.


When the coupler is on, the cavity field should couple strongly to the output port, characterized by the maximal coupling rate $\kappa_{\text{max}}$. 
But in order to detect if new loss channels are introduced by varying the bias, we extract this maximal coupling rate from the measured decay rate $\kappa$ in the time domain rather than the frequency domain.
The protocol (as depicted in the inset) is to drive the cavity at its resonance frequency with the coupler on, so that the cavity field reaches a steady state, corresponding to approximately $E_{\text{stored}} = 8000$ photons (Fig.~\ref{fig:2}b). Then, both the drive and the coupler are turned off for 1~$\mu$s to allow transients to settle. Finally the coupler is turned back on at varying bias and the outgoing field is measured. 
Because the energy in the cavity immediately before we turn on the varying bias is kept constant, we can measure the portion of this energy that reaches the output port and therefore identify additional loss channels.
Figure~2b shows three examples of time traces of the outgoing microwave field amplitude at different biases $\Phi/\Phi_0$.
Figure~2c shows the decay rate as a function of bias extracted from this procedure and the measured energy of the outgoing field.
When the bias is tuned from under-coupled to over-coupled, the external coupling becomes the dominant source of loss and the energy plateaus. But at even higher bias, we observe a reduction in the measured energy, so only the decay rate measured at the plateau can be interpreted as the achievable external coupling rate. As a result, we identify $\kappa_{\mathrm{max}}/2\pi = 1.96\ \pm 0.06$ MHz. 
We define the on/off ratio as $\kappa_{\mathrm{max}}/\kappa_{\mathrm{int}} =1130 \pm 40$, because the residual coupling when the coupler is off is much less than $\kappa_{\mathrm{int}}$ (see supplementary material). 

\begin{figure}
	\includegraphics[width = 246pt]{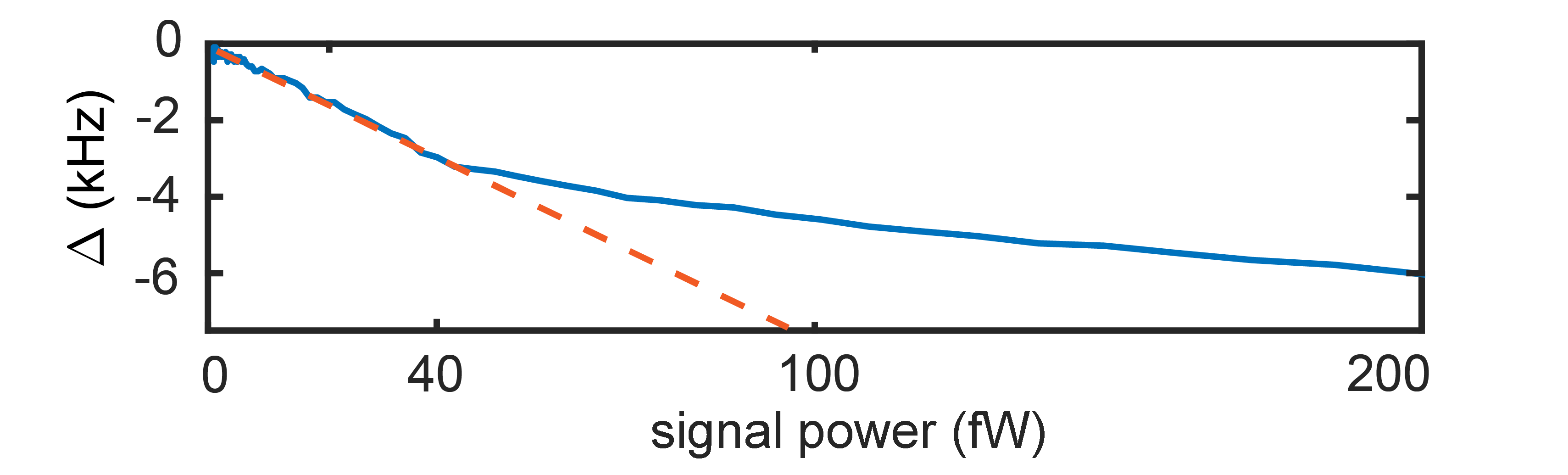}
	\caption{
    Measurements of coupler induced self-Kerr.
    We measure the cavity frequency shift from its low power limit $\Delta$ at different signal power (blue). A fit on the linear region is shown as the dashed orange line.
    At higher power, the resonator becomes bistable, deviating from the linear behavior.}
	\label{fig:3}
\end{figure}

To empty the cavity at the characteristic time of $1/\kappa_{\text{max}}$, it must also be possible to turn the coupler on/off much faster than that.
Fitting the time domain traces in Fig.~\ref{fig:2}b, we find that a lower bound on the switching speed is determined by the 50 MHz analog bandwidth of our analog-to-digital converter.
We model the time domain traces as decaying exponential pulses that have been modified by a low pass filter with corner frequency $\gamma_c/2 \pi$, and fit to the function:
\[    V(t) \propto
    \Theta(t - t_0) \frac{e^{-\gamma_c (t-t_0)} - e^{-\kappa (t-t_0)}}{\gamma_c - \kappa}\]
where $t_0$ is the start time, $\kappa$ is the total decay rate, and $\Theta(t-t_0)$ is the unit step function.
Although the rising edges in figure~2b appear instantaneous, they contain two data points. 
Here we focus on datasets at under-coupled biases, where the time traces are approximately step functions, to reduce the fit sensitivity of $\kappa$ and improve that of $\gamma_c$. 
We find $\gamma_c = 2\pi \times (48 \pm 5$~MHz) $\gg \kappa_{\text{max}}$, indicating that the switching time is at most $3.2 \pm 0.4$~ns.


 Finally, the coupler's ability to isolate the cavity from other microwave signals that might be present if this device were embedded in larger microwave network is also characterized by the cavity frequency shift induced either by an input signal from the measurement port or photons stored in the cavity. 
We characterize this self-Kerr effect by measuring $\Gamma$ and extracting the cavity frequency shift $\Delta$, caused by increasing the power in the probe tone above its low power limit. 
We determine the resonance frequency in these high-power reflection measurements as that frequency for which $\partial \angle \Gamma(\omega)/\partial \omega$ is maximum. 
The result is shown in figure~\ref{fig:3}.
These data are measured with the coupler bias tuned to critical coupling such that the number of photons $n$ in the cavity is $n = P_{\text{in}}/(\kappa \hbar\omega)$\rv{, where $P_{\text{in}}$ is the incident power at the output port, calibrated by dividing the attenuation of the lines from the microwave power injected into the cryostat}. Thus the Kerr shift can also be expressed in terms of the number of cavity photons as $-0.04 \pm 0.02$~Hz/photon. 
For operations using 1000 photons, the shift in the resonant frequency is still small compared to a linewidth. This compares favorably to other dynamic coupling implementations, which generally have $> 1$ kHz/photon self-Kerr \cite{Luke_2021, Pfaff_2017}.

\begin{table}
    \caption{\label{tab:table}Performance summary.}
        \begin{ruledtabular}
            \begin{tabular}{c|c}
            \hline
            loss and residual coupling              & 1.28 $\pm$ 0.01 kHz \\
            maximal coupling                        & 1.96\ $\pm$ 0.06 MHz\\
            on/off ratio                            & 1130 $\pm$ 40\\
            switching time                          & < 3.2 $\pm$ 0.4 ns \\
            self-Kerr                               & -0.04 $\pm$ 0.02  Hz/photon\\
            \end{tabular}
    \end{ruledtabular}
\end{table}

In conclusion, we have demonstrated an interface integrating a loop antenna and a coupling element that connects to a cavity, which has several desirable properties as building blocks for modular quantum networks. To that end, we envision extending this work by integrating multiple coupling elements onto one chip and creating a small quantum network with a few cavities, made monolithically, and connected by inserting such a chip into all of them simultaneously. We anticipate that desirable features demonstrated in this work can be preserved in such a scheme, indicating progress towards a larger, modular quantum network.

See the supplementary material for the device photo and wiring diagram, analysis on the coupling scheme, loop antenna design, uniform bias tuning steps and transmission measurements.

This work was supported by Q-SEnSE: Quantum Systems through Entangled Science and Engineering (NSF QLCI Award OMA-2016244) and the NSF Physics Frontier Center at JILA (Grant No. PHY-1734006).

The data that support the findings of this study are available from the corresponding author upon reasonable request.


\nocite{*}
\bibliography{bib}

\begin{thebibliography}{31}%
\makeatletter
\providecommand \@ifxundefined [1]{%
 \@ifx{#1\undefined}
}%
\providecommand \@ifnum [1]{%
 \ifnum #1\expandafter \@firstoftwo
 \else \expandafter \@secondoftwo
 \fi
}%
\providecommand \@ifx [1]{%
 \ifx #1\expandafter \@firstoftwo
 \else \expandafter \@secondoftwo
 \fi
}%
\providecommand \natexlab [1]{#1}%
\providecommand \enquote  [1]{``#1''}%
\providecommand \bibnamefont  [1]{#1}%
\providecommand \bibfnamefont [1]{#1}%
\providecommand \citenamefont [1]{#1}%
\providecommand \href@noop [0]{\@secondoftwo}%
\providecommand \href [0]{\begingroup \@sanitize@url \@href}%
\providecommand \@href[1]{\@@startlink{#1}\@@href}%
\providecommand \@@href[1]{\endgroup#1\@@endlink}%
\providecommand \@sanitize@url [0]{\catcode `\\12\catcode `\$12\catcode
  `\&12\catcode `\#12\catcode `\^12\catcode `\_12\catcode `\%12\relax}%
\providecommand \@@startlink[1]{}%
\providecommand \@@endlink[0]{}%
\providecommand \url  [0]{\begingroup\@sanitize@url \@url }%
\providecommand \@url [1]{\endgroup\@href {#1}{\urlprefix }}%
\providecommand \urlprefix  [0]{URL }%
\providecommand \Eprint [0]{\href }%
\providecommand \doibase [0]{http://dx.doi.org/}%
\providecommand \selectlanguage [0]{\@gobble}%
\providecommand \bibinfo  [0]{\@secondoftwo}%
\providecommand \bibfield  [0]{\@secondoftwo}%
\providecommand \translation [1]{[#1]}%
\providecommand \BibitemOpen [0]{}%
\providecommand \bibitemStop [0]{}%
\providecommand \bibitemNoStop [0]{.\EOS\space}%
\providecommand \EOS [0]{\spacefactor3000\relax}%
\providecommand \BibitemShut  [1]{\csname bibitem#1\endcsname}%
\let\auto@bib@innerbib\@empty
\bibitem [{\citenamefont {Brecht}\ \emph {et~al.}(2016)\citenamefont {Brecht},
  \citenamefont {Pfaff}, \citenamefont {Wang}, \citenamefont {Chu},
  \citenamefont {Frunzio}, \citenamefont {Devoret},\ and\ \citenamefont
  {Schoelkopf}}]{multilayer}%
  \BibitemOpen
  \bibfield  {author} {\bibinfo {author} {\bibfnamefont {T.}~\bibnamefont
  {Brecht}}, \bibinfo {author} {\bibfnamefont {W.}~\bibnamefont {Pfaff}},
  \bibinfo {author} {\bibfnamefont {C.}~\bibnamefont {Wang}}, \bibinfo {author}
  {\bibfnamefont {Y.}~\bibnamefont {Chu}}, \bibinfo {author} {\bibfnamefont
  {L.}~\bibnamefont {Frunzio}}, \bibinfo {author} {\bibfnamefont {M.~H.}\
  \bibnamefont {Devoret}}, \ and\ \bibinfo {author} {\bibfnamefont {R.~J.}\
  \bibnamefont {Schoelkopf}},\ }\href {\doibase 10.1038/npjqi.2016.2}
  {\bibfield  {journal} {\bibinfo  {journal} {npj Quantum Information}\
  }\textbf {\bibinfo {volume} {2}},\ \bibinfo {pages} {16002} (\bibinfo {year}
  {2016})}\BibitemShut {NoStop}%
\bibitem [{\citenamefont {Chou}\ \emph {et~al.}(2018)\citenamefont {Chou},
  \citenamefont {Blumoff}, \citenamefont {Wang}, \citenamefont {Reinhold},
  \citenamefont {Axline}, \citenamefont {Gao}, \citenamefont {Frunzio},
  \citenamefont {Devoret}, \citenamefont {Jiang},\ and\ \citenamefont
  {Schoelkopf}}]{Chou_2018}%
  \BibitemOpen
  \bibfield  {author} {\bibinfo {author} {\bibfnamefont {K.~S.}\ \bibnamefont
  {Chou}}, \bibinfo {author} {\bibfnamefont {J.~Z.}\ \bibnamefont {Blumoff}},
  \bibinfo {author} {\bibfnamefont {C.~S.}\ \bibnamefont {Wang}}, \bibinfo
  {author} {\bibfnamefont {P.~C.}\ \bibnamefont {Reinhold}}, \bibinfo {author}
  {\bibfnamefont {C.~J.}\ \bibnamefont {Axline}}, \bibinfo {author}
  {\bibfnamefont {Y.~Y.}\ \bibnamefont {Gao}}, \bibinfo {author} {\bibfnamefont
  {L.}~\bibnamefont {Frunzio}}, \bibinfo {author} {\bibfnamefont {M.~H.}\
  \bibnamefont {Devoret}}, \bibinfo {author} {\bibfnamefont {L.}~\bibnamefont
  {Jiang}}, \ and\ \bibinfo {author} {\bibfnamefont {R.~J.}\ \bibnamefont
  {Schoelkopf}},\ }\href {\doibase 10.1038/s41586-018-0470-y} {\bibfield
  {journal} {\bibinfo  {journal} {Nature}\ }\textbf {\bibinfo {volume} {561}},\
  \bibinfo {pages} {368} (\bibinfo {year} {2018})}\BibitemShut {NoStop}%
\bibitem [{\citenamefont {Zhou}\ \emph {et~al.}(2021)\citenamefont {Zhou},
  \citenamefont {Lu}, \citenamefont {Praquin}, \citenamefont {Chien},
  \citenamefont {Kaufman}, \citenamefont {Cao}, \citenamefont {Xia},
  \citenamefont {Mong}, \citenamefont {Pfaff}, \citenamefont {Pekker},\ and\
  \citenamefont {Hatridge}}]{Zhou_2021}%
  \BibitemOpen
  \bibfield  {author} {\bibinfo {author} {\bibfnamefont {C.}~\bibnamefont
  {Zhou}}, \bibinfo {author} {\bibfnamefont {P.}~\bibnamefont {Lu}}, \bibinfo
  {author} {\bibfnamefont {M.}~\bibnamefont {Praquin}}, \bibinfo {author}
  {\bibfnamefont {T.-C.}\ \bibnamefont {Chien}}, \bibinfo {author}
  {\bibfnamefont {R.}~\bibnamefont {Kaufman}}, \bibinfo {author} {\bibfnamefont
  {X.}~\bibnamefont {Cao}}, \bibinfo {author} {\bibfnamefont {M.}~\bibnamefont
  {Xia}}, \bibinfo {author} {\bibfnamefont {R.}~\bibnamefont {Mong}}, \bibinfo
  {author} {\bibfnamefont {W.}~\bibnamefont {Pfaff}}, \bibinfo {author}
  {\bibfnamefont {D.}~\bibnamefont {Pekker}}, \ and\ \bibinfo {author}
  {\bibfnamefont {M.}~\bibnamefont {Hatridge}},\ }\href {\doibase
  10.48550/ARXIV.2109.06848} {\enquote {\bibinfo {title} {A modular quantum
  computer based on a quantum state router},}\ } (\bibinfo {year}
  {2021})\BibitemShut {NoStop}%
\bibitem [{\citenamefont {Magnard}\ \emph {et~al.}(2020)\citenamefont
  {Magnard}, \citenamefont {Storz}, \citenamefont {Kurpiers}, \citenamefont
  {Sch\"ar}, \citenamefont {Marxer}, \citenamefont {L\"utolf}, \citenamefont
  {Walter}, \citenamefont {Besse}, \citenamefont {Gabureac}, \citenamefont
  {Reuer}, \citenamefont {Akin}, \citenamefont {Royer}, \citenamefont {Blais},\
  and\ \citenamefont {Wallraff}}]{Magnard_2020}%
  \BibitemOpen
  \bibfield  {author} {\bibinfo {author} {\bibfnamefont {P.}~\bibnamefont
  {Magnard}}, \bibinfo {author} {\bibfnamefont {S.}~\bibnamefont {Storz}},
  \bibinfo {author} {\bibfnamefont {P.}~\bibnamefont {Kurpiers}}, \bibinfo
  {author} {\bibfnamefont {J.}~\bibnamefont {Sch\"ar}}, \bibinfo {author}
  {\bibfnamefont {F.}~\bibnamefont {Marxer}}, \bibinfo {author} {\bibfnamefont
  {J.}~\bibnamefont {L\"utolf}}, \bibinfo {author} {\bibfnamefont
  {T.}~\bibnamefont {Walter}}, \bibinfo {author} {\bibfnamefont {J.-C.}\
  \bibnamefont {Besse}}, \bibinfo {author} {\bibfnamefont {M.}~\bibnamefont
  {Gabureac}}, \bibinfo {author} {\bibfnamefont {K.}~\bibnamefont {Reuer}},
  \bibinfo {author} {\bibfnamefont {A.}~\bibnamefont {Akin}}, \bibinfo {author}
  {\bibfnamefont {B.}~\bibnamefont {Royer}}, \bibinfo {author} {\bibfnamefont
  {A.}~\bibnamefont {Blais}}, \ and\ \bibinfo {author} {\bibfnamefont
  {A.}~\bibnamefont {Wallraff}},\ }\href {\doibase
  10.1103/PhysRevLett.125.260502} {\bibfield  {journal} {\bibinfo  {journal}
  {Phys. Rev. Lett.}\ }\textbf {\bibinfo {volume} {125}},\ \bibinfo {pages}
  {260502} (\bibinfo {year} {2020})}\BibitemShut {NoStop}%
\bibitem [{\citenamefont {Niu}\ \emph {et~al.}(2023)\citenamefont {Niu},
  \citenamefont {Zhang}, \citenamefont {Liu}, \citenamefont {Qiu},
  \citenamefont {Huang}, \citenamefont {Huang}, \citenamefont {Jia},
  \citenamefont {Liu}, \citenamefont {Tao}, \citenamefont {Wei}, \citenamefont
  {Zhou}, \citenamefont {Zou}, \citenamefont {Chen}, \citenamefont {Deng},
  \citenamefont {Deng}, \citenamefont {Hu}, \citenamefont {Hu}, \citenamefont
  {Li}, \citenamefont {Tan}, \citenamefont {Xu}, \citenamefont {Yan},
  \citenamefont {Yan}, \citenamefont {Liu}, \citenamefont {Zhong},
  \citenamefont {Cleland},\ and\ \citenamefont {Yu}}]{niu_low-loss_2023}%
  \BibitemOpen
  \bibfield  {author} {\bibinfo {author} {\bibfnamefont {J.}~\bibnamefont
  {Niu}}, \bibinfo {author} {\bibfnamefont {L.}~\bibnamefont {Zhang}}, \bibinfo
  {author} {\bibfnamefont {Y.}~\bibnamefont {Liu}}, \bibinfo {author}
  {\bibfnamefont {J.}~\bibnamefont {Qiu}}, \bibinfo {author} {\bibfnamefont
  {W.}~\bibnamefont {Huang}}, \bibinfo {author} {\bibfnamefont
  {J.}~\bibnamefont {Huang}}, \bibinfo {author} {\bibfnamefont
  {H.}~\bibnamefont {Jia}}, \bibinfo {author} {\bibfnamefont {J.}~\bibnamefont
  {Liu}}, \bibinfo {author} {\bibfnamefont {Z.}~\bibnamefont {Tao}}, \bibinfo
  {author} {\bibfnamefont {W.}~\bibnamefont {Wei}}, \bibinfo {author}
  {\bibfnamefont {Y.}~\bibnamefont {Zhou}}, \bibinfo {author} {\bibfnamefont
  {W.}~\bibnamefont {Zou}}, \bibinfo {author} {\bibfnamefont {Y.}~\bibnamefont
  {Chen}}, \bibinfo {author} {\bibfnamefont {X.}~\bibnamefont {Deng}}, \bibinfo
  {author} {\bibfnamefont {X.}~\bibnamefont {Deng}}, \bibinfo {author}
  {\bibfnamefont {C.}~\bibnamefont {Hu}}, \bibinfo {author} {\bibfnamefont
  {L.}~\bibnamefont {Hu}}, \bibinfo {author} {\bibfnamefont {J.}~\bibnamefont
  {Li}}, \bibinfo {author} {\bibfnamefont {D.}~\bibnamefont {Tan}}, \bibinfo
  {author} {\bibfnamefont {Y.}~\bibnamefont {Xu}}, \bibinfo {author}
  {\bibfnamefont {F.}~\bibnamefont {Yan}}, \bibinfo {author} {\bibfnamefont
  {T.}~\bibnamefont {Yan}}, \bibinfo {author} {\bibfnamefont {S.}~\bibnamefont
  {Liu}}, \bibinfo {author} {\bibfnamefont {Y.}~\bibnamefont {Zhong}}, \bibinfo
  {author} {\bibfnamefont {A.~N.}\ \bibnamefont {Cleland}}, \ and\ \bibinfo
  {author} {\bibfnamefont {D.}~\bibnamefont {Yu}},\ }\href {\doibase
  10.1038/s41928-023-00925-z} {\bibfield  {journal} {\bibinfo  {journal}
  {Nature Electronics}\ }\textbf {\bibinfo {volume} {6}},\ \bibinfo {pages}
  {235} (\bibinfo {year} {2023})}\BibitemShut {NoStop}%
\bibitem [{\citenamefont {Kannan}\ \emph {et~al.}(2023)\citenamefont {Kannan},
  \citenamefont {Almanakly}, \citenamefont {Sung}, \citenamefont {Di~Paolo},
  \citenamefont {Rower}, \citenamefont {Braumüller}, \citenamefont {Melville},
  \citenamefont {Niedzielski}, \citenamefont {Karamlou}, \citenamefont
  {Serniak}, \citenamefont {Vepsäläinen}, \citenamefont {Schwartz},
  \citenamefont {Yoder}, \citenamefont {Winik}, \citenamefont {Wang},
  \citenamefont {Orlando}, \citenamefont {Gustavsson}, \citenamefont {Grover},\
  and\ \citenamefont {Oliver}}]{kannan_waveguide_2023}%
  \BibitemOpen
  \bibfield  {author} {\bibinfo {author} {\bibfnamefont {B.}~\bibnamefont
  {Kannan}}, \bibinfo {author} {\bibfnamefont {A.}~\bibnamefont {Almanakly}},
  \bibinfo {author} {\bibfnamefont {Y.}~\bibnamefont {Sung}}, \bibinfo {author}
  {\bibfnamefont {A.}~\bibnamefont {Di~Paolo}}, \bibinfo {author}
  {\bibfnamefont {D.~A.}\ \bibnamefont {Rower}}, \bibinfo {author}
  {\bibfnamefont {J.}~\bibnamefont {Braumüller}}, \bibinfo {author}
  {\bibfnamefont {A.}~\bibnamefont {Melville}}, \bibinfo {author}
  {\bibfnamefont {B.~M.}\ \bibnamefont {Niedzielski}}, \bibinfo {author}
  {\bibfnamefont {A.}~\bibnamefont {Karamlou}}, \bibinfo {author}
  {\bibfnamefont {K.}~\bibnamefont {Serniak}}, \bibinfo {author} {\bibfnamefont
  {A.}~\bibnamefont {Vepsäläinen}}, \bibinfo {author} {\bibfnamefont {M.~E.}\
  \bibnamefont {Schwartz}}, \bibinfo {author} {\bibfnamefont {J.~L.}\
  \bibnamefont {Yoder}}, \bibinfo {author} {\bibfnamefont {R.}~\bibnamefont
  {Winik}}, \bibinfo {author} {\bibfnamefont {J.~I.-J.}\ \bibnamefont {Wang}},
  \bibinfo {author} {\bibfnamefont {T.~P.}\ \bibnamefont {Orlando}}, \bibinfo
  {author} {\bibfnamefont {S.}~\bibnamefont {Gustavsson}}, \bibinfo {author}
  {\bibfnamefont {J.~A.}\ \bibnamefont {Grover}}, \ and\ \bibinfo {author}
  {\bibfnamefont {W.~D.}\ \bibnamefont {Oliver}},\ }\href {\doibase
  10.1038/s41567-022-01869-5} {\bibfield  {journal} {\bibinfo  {journal}
  {Nature Physics}\ }\textbf {\bibinfo {volume} {19}},\ \bibinfo {pages} {394}
  (\bibinfo {year} {2023})}\BibitemShut {NoStop}%
\bibitem [{\citenamefont {Wang}\ \emph {et~al.}(2016)\citenamefont {Wang},
  \citenamefont {Gao}, \citenamefont {Reinhold}, \citenamefont {Heeres},
  \citenamefont {Ofek}, \citenamefont {Chou}, \citenamefont {Axline},
  \citenamefont {Reagor}, \citenamefont {Blumoff}, \citenamefont {Sliwa},
  \citenamefont {Frunzio}, \citenamefont {Girvin}, \citenamefont {Jiang},
  \citenamefont {Mirrahimi}, \citenamefont {Devoret},\ and\ \citenamefont
  {Schoelkopf}}]{Wang_2016}%
  \BibitemOpen
  \bibfield  {author} {\bibinfo {author} {\bibfnamefont {C.}~\bibnamefont
  {Wang}}, \bibinfo {author} {\bibfnamefont {Y.~Y.}\ \bibnamefont {Gao}},
  \bibinfo {author} {\bibfnamefont {P.}~\bibnamefont {Reinhold}}, \bibinfo
  {author} {\bibfnamefont {R.~W.}\ \bibnamefont {Heeres}}, \bibinfo {author}
  {\bibfnamefont {N.}~\bibnamefont {Ofek}}, \bibinfo {author} {\bibfnamefont
  {K.}~\bibnamefont {Chou}}, \bibinfo {author} {\bibfnamefont {C.}~\bibnamefont
  {Axline}}, \bibinfo {author} {\bibfnamefont {M.}~\bibnamefont {Reagor}},
  \bibinfo {author} {\bibfnamefont {J.}~\bibnamefont {Blumoff}}, \bibinfo
  {author} {\bibfnamefont {K.~M.}\ \bibnamefont {Sliwa}}, \bibinfo {author}
  {\bibfnamefont {L.}~\bibnamefont {Frunzio}}, \bibinfo {author} {\bibfnamefont
  {S.~M.}\ \bibnamefont {Girvin}}, \bibinfo {author} {\bibfnamefont
  {L.}~\bibnamefont {Jiang}}, \bibinfo {author} {\bibfnamefont
  {M.}~\bibnamefont {Mirrahimi}}, \bibinfo {author} {\bibfnamefont {M.~H.}\
  \bibnamefont {Devoret}}, \ and\ \bibinfo {author} {\bibfnamefont {R.~J.}\
  \bibnamefont {Schoelkopf}},\ }\href {\doibase 10.1126/science.aaf2941}
  {\bibfield  {journal} {\bibinfo  {journal} {Science}\ }\textbf {\bibinfo
  {volume} {352}},\ \bibinfo {pages} {1087} (\bibinfo {year} {2016})},\ \Eprint
  {http://arxiv.org/abs/https://www.science.org/doi/pdf/10.1126/science.aaf2941}
  {https://www.science.org/doi/pdf/10.1126/science.aaf2941} \BibitemShut
  {NoStop}%
\bibitem [{\citenamefont {Reagor}\ \emph {et~al.}(2016)\citenamefont {Reagor},
  \citenamefont {Pfaff}, \citenamefont {Axline}, \citenamefont {Heeres},
  \citenamefont {Ofek}, \citenamefont {Sliwa}, \citenamefont {Holland},
  \citenamefont {Wang}, \citenamefont {Blumoff}, \citenamefont {Chou},
  \citenamefont {Hatridge}, \citenamefont {Frunzio}, \citenamefont {Devoret},
  \citenamefont {Jiang},\ and\ \citenamefont {Schoelkopf}}]{Reagor_2016}%
  \BibitemOpen
  \bibfield  {author} {\bibinfo {author} {\bibfnamefont {M.}~\bibnamefont
  {Reagor}}, \bibinfo {author} {\bibfnamefont {W.}~\bibnamefont {Pfaff}},
  \bibinfo {author} {\bibfnamefont {C.}~\bibnamefont {Axline}}, \bibinfo
  {author} {\bibfnamefont {R.~W.}\ \bibnamefont {Heeres}}, \bibinfo {author}
  {\bibfnamefont {N.}~\bibnamefont {Ofek}}, \bibinfo {author} {\bibfnamefont
  {K.}~\bibnamefont {Sliwa}}, \bibinfo {author} {\bibfnamefont
  {E.}~\bibnamefont {Holland}}, \bibinfo {author} {\bibfnamefont
  {C.}~\bibnamefont {Wang}}, \bibinfo {author} {\bibfnamefont {J.}~\bibnamefont
  {Blumoff}}, \bibinfo {author} {\bibfnamefont {K.}~\bibnamefont {Chou}},
  \bibinfo {author} {\bibfnamefont {M.~J.}\ \bibnamefont {Hatridge}}, \bibinfo
  {author} {\bibfnamefont {L.}~\bibnamefont {Frunzio}}, \bibinfo {author}
  {\bibfnamefont {M.~H.}\ \bibnamefont {Devoret}}, \bibinfo {author}
  {\bibfnamefont {L.}~\bibnamefont {Jiang}}, \ and\ \bibinfo {author}
  {\bibfnamefont {R.~J.}\ \bibnamefont {Schoelkopf}},\ }\href {\doibase
  10.1103/PhysRevB.94.014506} {\bibfield  {journal} {\bibinfo  {journal} {Phys.
  Rev. B}\ }\textbf {\bibinfo {volume} {94}},\ \bibinfo {pages} {014506}
  (\bibinfo {year} {2016})}\BibitemShut {NoStop}%
\bibitem [{\citenamefont {Chakram}\ \emph {et~al.}(2021)\citenamefont
  {Chakram}, \citenamefont {Oriani}, \citenamefont {Naik}, \citenamefont
  {Dixit}, \citenamefont {He}, \citenamefont {Agrawal}, \citenamefont {Kwon},\
  and\ \citenamefont {Schuster}}]{Chakram_2021}%
  \BibitemOpen
  \bibfield  {author} {\bibinfo {author} {\bibfnamefont {S.}~\bibnamefont
  {Chakram}}, \bibinfo {author} {\bibfnamefont {A.~E.}\ \bibnamefont {Oriani}},
  \bibinfo {author} {\bibfnamefont {R.~K.}\ \bibnamefont {Naik}}, \bibinfo
  {author} {\bibfnamefont {A.~V.}\ \bibnamefont {Dixit}}, \bibinfo {author}
  {\bibfnamefont {K.}~\bibnamefont {He}}, \bibinfo {author} {\bibfnamefont
  {A.}~\bibnamefont {Agrawal}}, \bibinfo {author} {\bibfnamefont
  {H.}~\bibnamefont {Kwon}}, \ and\ \bibinfo {author} {\bibfnamefont {D.~I.}\
  \bibnamefont {Schuster}},\ }\href {\doibase 10.1103/PhysRevLett.127.107701}
  {\bibfield  {journal} {\bibinfo  {journal} {Phys. Rev. Lett.}\ }\textbf
  {\bibinfo {volume} {127}},\ \bibinfo {pages} {107701} (\bibinfo {year}
  {2021})}\BibitemShut {NoStop}%
\bibitem [{\citenamefont {Pfaff}\ \emph {et~al.}(2017)\citenamefont {Pfaff},
  \citenamefont {Axline}, \citenamefont {Burkhart}, \citenamefont {Vool},
  \citenamefont {Reinhold}, \citenamefont {Frunzio}, \citenamefont {Jiang},
  \citenamefont {Devoret},\ and\ \citenamefont {Schoelkopf}}]{Pfaff_2017}%
  \BibitemOpen
  \bibfield  {author} {\bibinfo {author} {\bibfnamefont {W.}~\bibnamefont
  {Pfaff}}, \bibinfo {author} {\bibfnamefont {C.~J.}\ \bibnamefont {Axline}},
  \bibinfo {author} {\bibfnamefont {L.~D.}\ \bibnamefont {Burkhart}}, \bibinfo
  {author} {\bibfnamefont {U.}~\bibnamefont {Vool}}, \bibinfo {author}
  {\bibfnamefont {P.}~\bibnamefont {Reinhold}}, \bibinfo {author}
  {\bibfnamefont {L.}~\bibnamefont {Frunzio}}, \bibinfo {author} {\bibfnamefont
  {L.}~\bibnamefont {Jiang}}, \bibinfo {author} {\bibfnamefont {M.~H.}\
  \bibnamefont {Devoret}}, \ and\ \bibinfo {author} {\bibfnamefont {R.~J.}\
  \bibnamefont {Schoelkopf}},\ }\href {\doibase 10.1038/nphys4143} {\bibfield
  {journal} {\bibinfo  {journal} {Nature Physics}\ }\textbf {\bibinfo {volume}
  {13}},\ \bibinfo {pages} {882} (\bibinfo {year} {2017})}\BibitemShut
  {NoStop}%
\bibitem [{\citenamefont {Burkhart}\ \emph {et~al.}(2021)\citenamefont
  {Burkhart}, \citenamefont {Teoh}, \citenamefont {Zhang}, \citenamefont
  {Axline}, \citenamefont {Frunzio}, \citenamefont {Devoret}, \citenamefont
  {Jiang}, \citenamefont {Girvin},\ and\ \citenamefont
  {Schoelkopf}}]{Luke_2021}%
  \BibitemOpen
  \bibfield  {author} {\bibinfo {author} {\bibfnamefont {L.~D.}\ \bibnamefont
  {Burkhart}}, \bibinfo {author} {\bibfnamefont {J.~D.}\ \bibnamefont {Teoh}},
  \bibinfo {author} {\bibfnamefont {Y.}~\bibnamefont {Zhang}}, \bibinfo
  {author} {\bibfnamefont {C.~J.}\ \bibnamefont {Axline}}, \bibinfo {author}
  {\bibfnamefont {L.}~\bibnamefont {Frunzio}}, \bibinfo {author} {\bibfnamefont
  {M.}~\bibnamefont {Devoret}}, \bibinfo {author} {\bibfnamefont
  {L.}~\bibnamefont {Jiang}}, \bibinfo {author} {\bibfnamefont
  {S.}~\bibnamefont {Girvin}}, \ and\ \bibinfo {author} {\bibfnamefont
  {R.}~\bibnamefont {Schoelkopf}},\ }\href {\doibase
  10.1103/PRXQuantum.2.030321} {\bibfield  {journal} {\bibinfo  {journal} {PRX
  Quantum}\ }\textbf {\bibinfo {volume} {2}},\ \bibinfo {pages} {030321}
  (\bibinfo {year} {2021})}\BibitemShut {NoStop}%
\bibitem [{\citenamefont {Axline}\ \emph {et~al.}(2018)\citenamefont {Axline},
  \citenamefont {Burkhart}, \citenamefont {Pfaff}, \citenamefont {Zhang},
  \citenamefont {Chou}, \citenamefont {Campagne-Ibarcq}, \citenamefont
  {Reinhold}, \citenamefont {Frunzio}, \citenamefont {Girvin}, \citenamefont
  {Jiang}, \citenamefont {Devoret},\ and\ \citenamefont
  {Schoelkopf}}]{Axline_2017}%
  \BibitemOpen
  \bibfield  {author} {\bibinfo {author} {\bibfnamefont {C.~J.}\ \bibnamefont
  {Axline}}, \bibinfo {author} {\bibfnamefont {L.~D.}\ \bibnamefont
  {Burkhart}}, \bibinfo {author} {\bibfnamefont {W.}~\bibnamefont {Pfaff}},
  \bibinfo {author} {\bibfnamefont {M.}~\bibnamefont {Zhang}}, \bibinfo
  {author} {\bibfnamefont {K.}~\bibnamefont {Chou}}, \bibinfo {author}
  {\bibfnamefont {P.}~\bibnamefont {Campagne-Ibarcq}}, \bibinfo {author}
  {\bibfnamefont {P.}~\bibnamefont {Reinhold}}, \bibinfo {author}
  {\bibfnamefont {L.}~\bibnamefont {Frunzio}}, \bibinfo {author} {\bibfnamefont
  {S.~M.}\ \bibnamefont {Girvin}}, \bibinfo {author} {\bibfnamefont
  {L.}~\bibnamefont {Jiang}}, \bibinfo {author} {\bibfnamefont {M.~H.}\
  \bibnamefont {Devoret}}, \ and\ \bibinfo {author} {\bibfnamefont {R.~J.}\
  \bibnamefont {Schoelkopf}},\ }\href {\doibase 10.1038/s41567-018-0115-y}
  {\bibfield  {journal} {\bibinfo  {journal} {Nature Physics}\ }\textbf
  {\bibinfo {volume} {14}},\ \bibinfo {pages} {705} (\bibinfo {year}
  {2018})}\BibitemShut {NoStop}%
\bibitem [{\citenamefont {Spring}\ \emph {et~al.}(2022)\citenamefont {Spring},
  \citenamefont {Cao}, \citenamefont {Tsunoda}, \citenamefont {Campanaro},
  \citenamefont {Fasciati}, \citenamefont {Wills}, \citenamefont {Bakr},
  \citenamefont {Chidambaram}, \citenamefont {Shteynas}, \citenamefont
  {Carpenter}, \citenamefont {Gow}, \citenamefont {Gates}, \citenamefont
  {Vlastakis},\ and\ \citenamefont {Leek}}]{Peter_2022}%
  \BibitemOpen
  \bibfield  {author} {\bibinfo {author} {\bibfnamefont {P.~A.}\ \bibnamefont
  {Spring}}, \bibinfo {author} {\bibfnamefont {S.}~\bibnamefont {Cao}},
  \bibinfo {author} {\bibfnamefont {T.}~\bibnamefont {Tsunoda}}, \bibinfo
  {author} {\bibfnamefont {G.}~\bibnamefont {Campanaro}}, \bibinfo {author}
  {\bibfnamefont {S.}~\bibnamefont {Fasciati}}, \bibinfo {author}
  {\bibfnamefont {J.}~\bibnamefont {Wills}}, \bibinfo {author} {\bibfnamefont
  {M.}~\bibnamefont {Bakr}}, \bibinfo {author} {\bibfnamefont {V.}~\bibnamefont
  {Chidambaram}}, \bibinfo {author} {\bibfnamefont {B.}~\bibnamefont
  {Shteynas}}, \bibinfo {author} {\bibfnamefont {L.}~\bibnamefont {Carpenter}},
  \bibinfo {author} {\bibfnamefont {P.}~\bibnamefont {Gow}}, \bibinfo {author}
  {\bibfnamefont {J.}~\bibnamefont {Gates}}, \bibinfo {author} {\bibfnamefont
  {B.}~\bibnamefont {Vlastakis}}, \ and\ \bibinfo {author} {\bibfnamefont
  {P.~J.}\ \bibnamefont {Leek}},\ }\href {\doibase 10.1126/sciadv.abl6698}
  {\bibfield  {journal} {\bibinfo  {journal} {Science Advances}\ }\textbf
  {\bibinfo {volume} {8}},\ \bibinfo {pages} {eabl6698} (\bibinfo {year}
  {2022})},\ \Eprint
  {http://arxiv.org/abs/https://www.science.org/doi/pdf/10.1126/sciadv.abl6698}
  {https://www.science.org/doi/pdf/10.1126/sciadv.abl6698} \BibitemShut
  {NoStop}%
\bibitem [{\citenamefont {Chapman}\ \emph {et~al.}(2016)\citenamefont
  {Chapman}, \citenamefont {Moores}, \citenamefont {Rosenthal}, \citenamefont
  {Kerckhoff},\ and\ \citenamefont {Lehnert}}]{TIB1}%
  \BibitemOpen
  \bibfield  {author} {\bibinfo {author} {\bibfnamefont {B.}~\bibnamefont
  {Chapman}}, \bibinfo {author} {\bibfnamefont {B.}~\bibnamefont {Moores}},
  \bibinfo {author} {\bibfnamefont {E.}~\bibnamefont {Rosenthal}}, \bibinfo
  {author} {\bibfnamefont {J.}~\bibnamefont {Kerckhoff}}, \ and\ \bibinfo
  {author} {\bibfnamefont {K.}~\bibnamefont {Lehnert}},\ }\href {\doibase
  10.1063/1.4952772} {\bibfield  {journal} {\bibinfo  {journal} {Applied
  Physics Letters}\ }\textbf {\bibinfo {volume} {108}},\ \bibinfo {pages}
  {222602} (\bibinfo {year} {2016})}\BibitemShut {NoStop}%
\bibitem [{\citenamefont {Brecht}\ \emph {et~al.}(2015)\citenamefont {Brecht},
  \citenamefont {Reagor}, \citenamefont {Chu}, \citenamefont {Pfaff},
  \citenamefont {Wang}, \citenamefont {Frunzio}, \citenamefont {Devoret},\ and\
  \citenamefont {Schoelkopf}}]{Demonstration_Brecht}%
  \BibitemOpen
  \bibfield  {author} {\bibinfo {author} {\bibfnamefont {T.}~\bibnamefont
  {Brecht}}, \bibinfo {author} {\bibfnamefont {M.}~\bibnamefont {Reagor}},
  \bibinfo {author} {\bibfnamefont {Y.}~\bibnamefont {Chu}}, \bibinfo {author}
  {\bibfnamefont {W.}~\bibnamefont {Pfaff}}, \bibinfo {author} {\bibfnamefont
  {C.}~\bibnamefont {Wang}}, \bibinfo {author} {\bibfnamefont {L.}~\bibnamefont
  {Frunzio}}, \bibinfo {author} {\bibfnamefont {M.~H.}\ \bibnamefont
  {Devoret}}, \ and\ \bibinfo {author} {\bibfnamefont {R.~J.}\ \bibnamefont
  {Schoelkopf}},\ }\href {\doibase 10.1063/1.4935541} {\bibfield  {journal}
  {\bibinfo  {journal} {Applied Physics Letters}\ }\textbf {\bibinfo {volume}
  {107}} (\bibinfo {year} {2015}),\ 10.1063/1.4935541},\ \bibinfo {note}
  {192603},\ \Eprint
  {http://arxiv.org/abs/https://pubs.aip.org/aip/apl/article-pdf/doi/10.1063/1.4935541/13656164/192603\_1\_online.pdf}
  {https://pubs.aip.org/aip/apl/article-pdf/doi/10.1063/1.4935541/13656164/192603\_1\_online.pdf}
  \BibitemShut {NoStop}%
\bibitem [{\citenamefont {Campagne-Ibarcq}\ \emph {et~al.}(2020)\citenamefont
  {Campagne-Ibarcq}, \citenamefont {Eickbusch}, \citenamefont {Touzard},
  \citenamefont {Zalys-Geller}, \citenamefont {Frattini}, \citenamefont
  {Sivak}, \citenamefont {Reinhold}, \citenamefont {Puri}, \citenamefont
  {Shankar}, \citenamefont {Schoelkopf}, \citenamefont {Frunzio}, \citenamefont
  {Mirrahimi},\ and\ \citenamefont {Devoret}}]{Campagne-Ibarcq2020}%
  \BibitemOpen
  \bibfield  {author} {\bibinfo {author} {\bibfnamefont {P.}~\bibnamefont
  {Campagne-Ibarcq}}, \bibinfo {author} {\bibfnamefont {A.}~\bibnamefont
  {Eickbusch}}, \bibinfo {author} {\bibfnamefont {S.}~\bibnamefont {Touzard}},
  \bibinfo {author} {\bibfnamefont {E.}~\bibnamefont {Zalys-Geller}}, \bibinfo
  {author} {\bibfnamefont {N.~E.}\ \bibnamefont {Frattini}}, \bibinfo {author}
  {\bibfnamefont {V.~V.}\ \bibnamefont {Sivak}}, \bibinfo {author}
  {\bibfnamefont {P.}~\bibnamefont {Reinhold}}, \bibinfo {author}
  {\bibfnamefont {S.}~\bibnamefont {Puri}}, \bibinfo {author} {\bibfnamefont
  {S.}~\bibnamefont {Shankar}}, \bibinfo {author} {\bibfnamefont {R.~J.}\
  \bibnamefont {Schoelkopf}}, \bibinfo {author} {\bibfnamefont
  {L.}~\bibnamefont {Frunzio}}, \bibinfo {author} {\bibfnamefont
  {M.}~\bibnamefont {Mirrahimi}}, \ and\ \bibinfo {author} {\bibfnamefont
  {M.~H.}\ \bibnamefont {Devoret}},\ }\href {\doibase
  10.1038/s41586-020-2603-3} {\bibfield  {journal} {\bibinfo  {journal}
  {Nature}\ }\textbf {\bibinfo {volume} {584}},\ \bibinfo {pages} {368}
  (\bibinfo {year} {2020})}\BibitemShut {NoStop}%
\bibitem [{\citenamefont {Mates}\ \emph {et~al.}(2008)\citenamefont {Mates},
  \citenamefont {Hilton}, \citenamefont {Irwin}, \citenamefont {Vale},\ and\
  \citenamefont {Lehnert}}]{Demonstration_Mates}%
  \BibitemOpen
  \bibfield  {author} {\bibinfo {author} {\bibfnamefont {J.~A.~B.}\
  \bibnamefont {Mates}}, \bibinfo {author} {\bibfnamefont {G.~C.}\ \bibnamefont
  {Hilton}}, \bibinfo {author} {\bibfnamefont {K.~D.}\ \bibnamefont {Irwin}},
  \bibinfo {author} {\bibfnamefont {L.~R.}\ \bibnamefont {Vale}}, \ and\
  \bibinfo {author} {\bibfnamefont {K.~W.}\ \bibnamefont {Lehnert}},\ }\href
  {\doibase 10.1063/1.2803852} {\bibfield  {journal} {\bibinfo  {journal}
  {Applied Physics Letters}\ }\textbf {\bibinfo {volume} {92}} (\bibinfo {year}
  {2008}),\ 10.1063/1.2803852},\ \bibinfo {note} {023514},\ \Eprint
  {http://arxiv.org/abs/https://pubs.aip.org/aip/apl/article-pdf/doi/10.1063/1.2803852/14385856/023514\_1\_online.pdf}
  {https://pubs.aip.org/aip/apl/article-pdf/doi/10.1063/1.2803852/14385856/023514\_1\_online.pdf}
  \BibitemShut {NoStop}%
\bibitem [{\citenamefont {Rosenthal}\ \emph {et~al.}(2021)\citenamefont
  {Rosenthal}, \citenamefont {Schneider}, \citenamefont {Malnou}, \citenamefont
  {Zhao}, \citenamefont {Leditzky}, \citenamefont {Chapman}, \citenamefont
  {Wustmann}, \citenamefont {Ma}, \citenamefont {Palken}, \citenamefont
  {Zanner}, \citenamefont {Vale}, \citenamefont {Hilton}, \citenamefont {Gao},
  \citenamefont {Smith}, \citenamefont {Kirchmair},\ and\ \citenamefont
  {Lehnert}}]{Eric_2021}%
  \BibitemOpen
  \bibfield  {author} {\bibinfo {author} {\bibfnamefont {E.~I.}\ \bibnamefont
  {Rosenthal}}, \bibinfo {author} {\bibfnamefont {C.~M.~F.}\ \bibnamefont
  {Schneider}}, \bibinfo {author} {\bibfnamefont {M.}~\bibnamefont {Malnou}},
  \bibinfo {author} {\bibfnamefont {Z.}~\bibnamefont {Zhao}}, \bibinfo {author}
  {\bibfnamefont {F.}~\bibnamefont {Leditzky}}, \bibinfo {author}
  {\bibfnamefont {B.~J.}\ \bibnamefont {Chapman}}, \bibinfo {author}
  {\bibfnamefont {W.}~\bibnamefont {Wustmann}}, \bibinfo {author}
  {\bibfnamefont {X.}~\bibnamefont {Ma}}, \bibinfo {author} {\bibfnamefont
  {D.~A.}\ \bibnamefont {Palken}}, \bibinfo {author} {\bibfnamefont {M.~F.}\
  \bibnamefont {Zanner}}, \bibinfo {author} {\bibfnamefont {L.~R.}\
  \bibnamefont {Vale}}, \bibinfo {author} {\bibfnamefont {G.~C.}\ \bibnamefont
  {Hilton}}, \bibinfo {author} {\bibfnamefont {J.}~\bibnamefont {Gao}},
  \bibinfo {author} {\bibfnamefont {G.}~\bibnamefont {Smith}}, \bibinfo
  {author} {\bibfnamefont {G.}~\bibnamefont {Kirchmair}}, \ and\ \bibinfo
  {author} {\bibfnamefont {K.~W.}\ \bibnamefont {Lehnert}},\ }\href {\doibase
  10.1103/PhysRevLett.126.090503} {\bibfield  {journal} {\bibinfo  {journal}
  {Phys. Rev. Lett.}\ }\textbf {\bibinfo {volume} {126}},\ \bibinfo {pages}
  {090503} (\bibinfo {year} {2021})}\BibitemShut {NoStop}%
\bibitem [{\citenamefont {Kerckhoff}\ \emph {et~al.}(2015)\citenamefont
  {Kerckhoff}, \citenamefont {Lalumi\`ere}, \citenamefont {Chapman},
  \citenamefont {Blais},\ and\ \citenamefont {Lehnert}}]{Kerckhoff_2015}%
  \BibitemOpen
  \bibfield  {author} {\bibinfo {author} {\bibfnamefont {J.}~\bibnamefont
  {Kerckhoff}}, \bibinfo {author} {\bibfnamefont {K.}~\bibnamefont
  {Lalumi\`ere}}, \bibinfo {author} {\bibfnamefont {B.~J.}\ \bibnamefont
  {Chapman}}, \bibinfo {author} {\bibfnamefont {A.}~\bibnamefont {Blais}}, \
  and\ \bibinfo {author} {\bibfnamefont {K.~W.}\ \bibnamefont {Lehnert}},\
  }\href {\doibase 10.1103/PhysRevApplied.4.034002} {\bibfield  {journal}
  {\bibinfo  {journal} {Phys. Rev. Appl.}\ }\textbf {\bibinfo {volume} {4}},\
  \bibinfo {pages} {034002} (\bibinfo {year} {2015})}\BibitemShut {NoStop}%
\bibitem [{\citenamefont {Bergeal}\ \emph {et~al.}(2010)\citenamefont
  {Bergeal}, \citenamefont {Vijay}, \citenamefont {Manucharyan}, \citenamefont
  {Siddiqi}, \citenamefont {Schoelkopf}, \citenamefont {Girvin},\ and\
  \citenamefont {Devoret}}]{Bergeal_JRM_2010}%
  \BibitemOpen
  \bibfield  {author} {\bibinfo {author} {\bibfnamefont {N.}~\bibnamefont
  {Bergeal}}, \bibinfo {author} {\bibfnamefont {R.}~\bibnamefont {Vijay}},
  \bibinfo {author} {\bibfnamefont {V.~E.}\ \bibnamefont {Manucharyan}},
  \bibinfo {author} {\bibfnamefont {I.}~\bibnamefont {Siddiqi}}, \bibinfo
  {author} {\bibfnamefont {R.~J.}\ \bibnamefont {Schoelkopf}}, \bibinfo
  {author} {\bibfnamefont {S.~M.}\ \bibnamefont {Girvin}}, \ and\ \bibinfo
  {author} {\bibfnamefont {M.~H.}\ \bibnamefont {Devoret}},\ }\href {\doibase
  10.1038/nphys1516} {\bibfield  {journal} {\bibinfo  {journal} {Nature
  Physics}\ }\textbf {\bibinfo {volume} {6}},\ \bibinfo {pages} {296} (\bibinfo
  {year} {2010})}\BibitemShut {NoStop}%
\bibitem [{\citenamefont {Flurin}\ \emph {et~al.}(2015)\citenamefont {Flurin},
  \citenamefont {Roch}, \citenamefont {Pillet}, \citenamefont {Mallet},\ and\
  \citenamefont {Huard}}]{Flurin_2015}%
  \BibitemOpen
  \bibfield  {author} {\bibinfo {author} {\bibfnamefont {E.}~\bibnamefont
  {Flurin}}, \bibinfo {author} {\bibfnamefont {N.}~\bibnamefont {Roch}},
  \bibinfo {author} {\bibfnamefont {J.~D.}\ \bibnamefont {Pillet}}, \bibinfo
  {author} {\bibfnamefont {F.}~\bibnamefont {Mallet}}, \ and\ \bibinfo {author}
  {\bibfnamefont {B.}~\bibnamefont {Huard}},\ }\href {\doibase
  10.1103/PhysRevLett.114.090503} {\bibfield  {journal} {\bibinfo  {journal}
  {Phys. Rev. Lett.}\ }\textbf {\bibinfo {volume} {114}},\ \bibinfo {pages}
  {090503} (\bibinfo {year} {2015})}\BibitemShut {NoStop}%
\bibitem [{\citenamefont {Gargiulo}\ \emph {et~al.}(2021)\citenamefont
  {Gargiulo}, \citenamefont {Oleschko}, \citenamefont {Prat-Camps},
  \citenamefont {Zanner},\ and\ \citenamefont {Kirchmair}}]{flux_hose}%
  \BibitemOpen
  \bibfield  {author} {\bibinfo {author} {\bibfnamefont {O.}~\bibnamefont
  {Gargiulo}}, \bibinfo {author} {\bibfnamefont {S.}~\bibnamefont {Oleschko}},
  \bibinfo {author} {\bibfnamefont {J.}~\bibnamefont {Prat-Camps}}, \bibinfo
  {author} {\bibfnamefont {M.}~\bibnamefont {Zanner}}, \ and\ \bibinfo {author}
  {\bibfnamefont {G.}~\bibnamefont {Kirchmair}},\ }\href {\doibase
  10.1063/5.0032615} {\bibfield  {journal} {\bibinfo  {journal} {Applied
  Physics Letters}\ }\textbf {\bibinfo {volume} {118}},\ \bibinfo {pages}
  {012601} (\bibinfo {year} {2021})},\ \Eprint
  {http://arxiv.org/abs/https://doi.org/10.1063/5.0032615}
  {https://doi.org/10.1063/5.0032615} \BibitemShut {NoStop}%
\bibitem [{\citenamefont {Gold}\ \emph {et~al.}(2021)\citenamefont {Gold},
  \citenamefont {Paquette}, \citenamefont {Stockklauser}, \citenamefont
  {Reagor}, \citenamefont {Alam}, \citenamefont {Bestwick}, \citenamefont
  {Didier}, \citenamefont {Nersisyan}, \citenamefont {Oruc}, \citenamefont
  {Razavi}, \citenamefont {Scharmann}, \citenamefont {Sete}, \citenamefont
  {Sur}, \citenamefont {Venturelli}, \citenamefont {Winkleblack}, \citenamefont
  {Wudarski}, \citenamefont {Harburn},\ and\ \citenamefont
  {Rigetti}}]{Gold_2021}%
  \BibitemOpen
  \bibfield  {author} {\bibinfo {author} {\bibfnamefont {A.}~\bibnamefont
  {Gold}}, \bibinfo {author} {\bibfnamefont {J.~P.}\ \bibnamefont {Paquette}},
  \bibinfo {author} {\bibfnamefont {A.}~\bibnamefont {Stockklauser}}, \bibinfo
  {author} {\bibfnamefont {M.~J.}\ \bibnamefont {Reagor}}, \bibinfo {author}
  {\bibfnamefont {M.~S.}\ \bibnamefont {Alam}}, \bibinfo {author}
  {\bibfnamefont {A.}~\bibnamefont {Bestwick}}, \bibinfo {author}
  {\bibfnamefont {N.}~\bibnamefont {Didier}}, \bibinfo {author} {\bibfnamefont
  {A.}~\bibnamefont {Nersisyan}}, \bibinfo {author} {\bibfnamefont
  {F.}~\bibnamefont {Oruc}}, \bibinfo {author} {\bibfnamefont {A.}~\bibnamefont
  {Razavi}}, \bibinfo {author} {\bibfnamefont {B.}~\bibnamefont {Scharmann}},
  \bibinfo {author} {\bibfnamefont {E.~A.}\ \bibnamefont {Sete}}, \bibinfo
  {author} {\bibfnamefont {B.}~\bibnamefont {Sur}}, \bibinfo {author}
  {\bibfnamefont {D.}~\bibnamefont {Venturelli}}, \bibinfo {author}
  {\bibfnamefont {C.~J.}\ \bibnamefont {Winkleblack}}, \bibinfo {author}
  {\bibfnamefont {F.}~\bibnamefont {Wudarski}}, \bibinfo {author}
  {\bibfnamefont {M.}~\bibnamefont {Harburn}}, \ and\ \bibinfo {author}
  {\bibfnamefont {C.}~\bibnamefont {Rigetti}},\ }\href {\doibase
  10.1038/s41534-021-00484-1} {\bibfield  {journal} {\bibinfo  {journal} {npj
  Quantum Information}\ }\textbf {\bibinfo {volume} {7}},\ \bibinfo {pages}
  {142} (\bibinfo {year} {2021})}\BibitemShut {NoStop}%
\bibitem [{\citenamefont {Lei}\ \emph {et~al.}(2020)\citenamefont {Lei},
  \citenamefont {Krayzman}, \citenamefont {Ganjam}, \citenamefont {Frunzio},\
  and\ \citenamefont {Schoelkopf}}]{Lei_2020}%
  \BibitemOpen
  \bibfield  {author} {\bibinfo {author} {\bibfnamefont {C.~U.}\ \bibnamefont
  {Lei}}, \bibinfo {author} {\bibfnamefont {L.}~\bibnamefont {Krayzman}},
  \bibinfo {author} {\bibfnamefont {S.}~\bibnamefont {Ganjam}}, \bibinfo
  {author} {\bibfnamefont {L.}~\bibnamefont {Frunzio}}, \ and\ \bibinfo
  {author} {\bibfnamefont {R.~J.}\ \bibnamefont {Schoelkopf}},\ }\href
  {\doibase 10.1063/5.0003907} {\bibfield  {journal} {\bibinfo  {journal}
  {Applied Physics Letters}\ }\textbf {\bibinfo {volume} {116}},\ \bibinfo
  {pages} {154002} (\bibinfo {year} {2020})},\ \Eprint
  {http://arxiv.org/abs/https://doi.org/10.1063/5.0003907}
  {https://doi.org/10.1063/5.0003907} \BibitemShut {NoStop}%
\bibitem [{\citenamefont {Paik}\ \emph {et~al.}(2011)\citenamefont {Paik},
  \citenamefont {Schuster}, \citenamefont {Bishop}, \citenamefont {Kirchmair},
  \citenamefont {Catelani}, \citenamefont {Sears}, \citenamefont {Johnson},
  \citenamefont {Reagor}, \citenamefont {Frunzio}, \citenamefont {Glazman},
  \citenamefont {Girvin}, \citenamefont {Devoret},\ and\ \citenamefont
  {Schoelkopf}}]{Paik_2011}%
  \BibitemOpen
  \bibfield  {author} {\bibinfo {author} {\bibfnamefont {H.}~\bibnamefont
  {Paik}}, \bibinfo {author} {\bibfnamefont {D.~I.}\ \bibnamefont {Schuster}},
  \bibinfo {author} {\bibfnamefont {L.~S.}\ \bibnamefont {Bishop}}, \bibinfo
  {author} {\bibfnamefont {G.}~\bibnamefont {Kirchmair}}, \bibinfo {author}
  {\bibfnamefont {G.}~\bibnamefont {Catelani}}, \bibinfo {author}
  {\bibfnamefont {A.~P.}\ \bibnamefont {Sears}}, \bibinfo {author}
  {\bibfnamefont {B.~R.}\ \bibnamefont {Johnson}}, \bibinfo {author}
  {\bibfnamefont {M.~J.}\ \bibnamefont {Reagor}}, \bibinfo {author}
  {\bibfnamefont {L.}~\bibnamefont {Frunzio}}, \bibinfo {author} {\bibfnamefont
  {L.~I.}\ \bibnamefont {Glazman}}, \bibinfo {author} {\bibfnamefont {S.~M.}\
  \bibnamefont {Girvin}}, \bibinfo {author} {\bibfnamefont {M.~H.}\
  \bibnamefont {Devoret}}, \ and\ \bibinfo {author} {\bibfnamefont {R.~J.}\
  \bibnamefont {Schoelkopf}},\ }\href {\doibase 10.1103/PhysRevLett.107.240501}
  {\bibfield  {journal} {\bibinfo  {journal} {Phys. Rev. Lett.}\ }\textbf
  {\bibinfo {volume} {107}},\ \bibinfo {pages} {240501} (\bibinfo {year}
  {2011})}\BibitemShut {NoStop}%
\bibitem [{\citenamefont {Lecocq}\ \emph {et~al.}(2021)\citenamefont {Lecocq},
  \citenamefont {Ranzani}, \citenamefont {Peterson}, \citenamefont {Cicak},
  \citenamefont {Jin}, \citenamefont {Simmonds}, \citenamefont {Teufel},\ and\
  \citenamefont {Aumentado}}]{FPJA_RO}%
  \BibitemOpen
  \bibfield  {author} {\bibinfo {author} {\bibfnamefont {F.}~\bibnamefont
  {Lecocq}}, \bibinfo {author} {\bibfnamefont {L.}~\bibnamefont {Ranzani}},
  \bibinfo {author} {\bibfnamefont {G.~A.}\ \bibnamefont {Peterson}}, \bibinfo
  {author} {\bibfnamefont {K.}~\bibnamefont {Cicak}}, \bibinfo {author}
  {\bibfnamefont {X.~Y.}\ \bibnamefont {Jin}}, \bibinfo {author} {\bibfnamefont
  {R.~W.}\ \bibnamefont {Simmonds}}, \bibinfo {author} {\bibfnamefont {J.~D.}\
  \bibnamefont {Teufel}}, \ and\ \bibinfo {author} {\bibfnamefont
  {J.}~\bibnamefont {Aumentado}},\ }\href {\doibase
  10.1103/PhysRevLett.126.020502} {\bibfield  {journal} {\bibinfo  {journal}
  {Phys. Rev. Lett.}\ }\textbf {\bibinfo {volume} {126}},\ \bibinfo {pages}
  {020502} (\bibinfo {year} {2021})}\BibitemShut {NoStop}%
\bibitem [{\citenamefont {Bialczak}\ \emph {et~al.}(2011)\citenamefont
  {Bialczak}, \citenamefont {Ansmann}, \citenamefont {Hofheinz}, \citenamefont
  {Lenander}, \citenamefont {Lucero}, \citenamefont {Neeley}, \citenamefont
  {O'Connell}, \citenamefont {Sank}, \citenamefont {Wang}, \citenamefont
  {Weides}, \citenamefont {Wenner}, \citenamefont {Yamamoto}, \citenamefont
  {Cleland},\ and\ \citenamefont {Martinis}}]{2011_coupler}%
  \BibitemOpen
  \bibfield  {author} {\bibinfo {author} {\bibfnamefont {R.~C.}\ \bibnamefont
  {Bialczak}}, \bibinfo {author} {\bibfnamefont {M.}~\bibnamefont {Ansmann}},
  \bibinfo {author} {\bibfnamefont {M.}~\bibnamefont {Hofheinz}}, \bibinfo
  {author} {\bibfnamefont {M.}~\bibnamefont {Lenander}}, \bibinfo {author}
  {\bibfnamefont {E.}~\bibnamefont {Lucero}}, \bibinfo {author} {\bibfnamefont
  {M.}~\bibnamefont {Neeley}}, \bibinfo {author} {\bibfnamefont {A.~D.}\
  \bibnamefont {O'Connell}}, \bibinfo {author} {\bibfnamefont {D.}~\bibnamefont
  {Sank}}, \bibinfo {author} {\bibfnamefont {H.}~\bibnamefont {Wang}}, \bibinfo
  {author} {\bibfnamefont {M.}~\bibnamefont {Weides}}, \bibinfo {author}
  {\bibfnamefont {J.}~\bibnamefont {Wenner}}, \bibinfo {author} {\bibfnamefont
  {T.}~\bibnamefont {Yamamoto}}, \bibinfo {author} {\bibfnamefont {A.~N.}\
  \bibnamefont {Cleland}}, \ and\ \bibinfo {author} {\bibfnamefont {J.~M.}\
  \bibnamefont {Martinis}},\ }\href {\doibase 10.1103/PhysRevLett.106.060501}
  {\bibfield  {journal} {\bibinfo  {journal} {Phys. Rev. Lett.}\ }\textbf
  {\bibinfo {volume} {106}},\ \bibinfo {pages} {060501} (\bibinfo {year}
  {2011})}\BibitemShut {NoStop}%
\bibitem [{\citenamefont {Pechal}\ \emph {et~al.}(2016)\citenamefont {Pechal},
  \citenamefont {Besse}, \citenamefont {Mondal}, \citenamefont {Oppliger},
  \citenamefont {Gasparinetti},\ and\ \citenamefont {Wallraff}}]{Pechal_2016}%
  \BibitemOpen
  \bibfield  {author} {\bibinfo {author} {\bibfnamefont {M.}~\bibnamefont
  {Pechal}}, \bibinfo {author} {\bibfnamefont {J.-C.}\ \bibnamefont {Besse}},
  \bibinfo {author} {\bibfnamefont {M.}~\bibnamefont {Mondal}}, \bibinfo
  {author} {\bibfnamefont {M.}~\bibnamefont {Oppliger}}, \bibinfo {author}
  {\bibfnamefont {S.}~\bibnamefont {Gasparinetti}}, \ and\ \bibinfo {author}
  {\bibfnamefont {A.}~\bibnamefont {Wallraff}},\ }\href {\doibase
  10.1103/PhysRevApplied.6.024009} {\bibfield  {journal} {\bibinfo  {journal}
  {Phys. Rev. Applied}\ }\textbf {\bibinfo {volume} {6}},\ \bibinfo {pages}
  {024009} (\bibinfo {year} {2016})}\BibitemShut {NoStop}%
\bibitem [{\citenamefont {Naaman}\ \emph {et~al.}(2016)\citenamefont {Naaman},
  \citenamefont {Abutaleb}, \citenamefont {Kirby},\ and\ \citenamefont
  {Rennie}}]{Naaman_2016}%
  \BibitemOpen
  \bibfield  {author} {\bibinfo {author} {\bibfnamefont {O.}~\bibnamefont
  {Naaman}}, \bibinfo {author} {\bibfnamefont {M.~O.}\ \bibnamefont
  {Abutaleb}}, \bibinfo {author} {\bibfnamefont {C.}~\bibnamefont {Kirby}}, \
  and\ \bibinfo {author} {\bibfnamefont {M.}~\bibnamefont {Rennie}},\
  }\bibfield  {booktitle} {\emph {\bibinfo {booktitle} {Applied Physics
  Letters}},\ }\href {\doibase 10.1063/1.4943602} {\bibfield  {journal}
  {\bibinfo  {journal} {Applied Physics Letters}\ }\textbf {\bibinfo {volume}
  {108}},\ \bibinfo {pages} {112601} (\bibinfo {year} {2016})}\BibitemShut
  {NoStop}%
\bibitem [{\citenamefont {Lecocq}\ \emph {et~al.}(2017)\citenamefont {Lecocq},
  \citenamefont {Ranzani}, \citenamefont {Peterson}, \citenamefont {Cicak},
  \citenamefont {Simmonds}, \citenamefont {Teufel},\ and\ \citenamefont
  {Aumentado}}]{Lacocq_2017}%
  \BibitemOpen
  \bibfield  {author} {\bibinfo {author} {\bibfnamefont {F.}~\bibnamefont
  {Lecocq}}, \bibinfo {author} {\bibfnamefont {L.}~\bibnamefont {Ranzani}},
  \bibinfo {author} {\bibfnamefont {G.~A.}\ \bibnamefont {Peterson}}, \bibinfo
  {author} {\bibfnamefont {K.}~\bibnamefont {Cicak}}, \bibinfo {author}
  {\bibfnamefont {R.~W.}\ \bibnamefont {Simmonds}}, \bibinfo {author}
  {\bibfnamefont {J.~D.}\ \bibnamefont {Teufel}}, \ and\ \bibinfo {author}
  {\bibfnamefont {J.}~\bibnamefont {Aumentado}},\ }\href {\doibase
  10.1103/PhysRevApplied.7.024028} {\bibfield  {journal} {\bibinfo  {journal}
  {Phys. Rev. Applied}\ }\textbf {\bibinfo {volume} {7}},\ \bibinfo {pages}
  {024028} (\bibinfo {year} {2017})}\BibitemShut {NoStop}%
\bibitem [{\citenamefont {Chang}\ \emph {et~al.}(2020)\citenamefont {Chang},
  \citenamefont {Satzinger}, \citenamefont {Zhong}, \citenamefont {Bienfait},
  \citenamefont {Chou}, \citenamefont {Conner}, \citenamefont {Dumur},
  \citenamefont {Grebel}, \citenamefont {Peairs}, \citenamefont {Povey},\ and\
  \citenamefont {Cleland}}]{Chang_2020}%
  \BibitemOpen
  \bibfield  {author} {\bibinfo {author} {\bibfnamefont {H.-S.}\ \bibnamefont
  {Chang}}, \bibinfo {author} {\bibfnamefont {K.~J.}\ \bibnamefont
  {Satzinger}}, \bibinfo {author} {\bibfnamefont {Y.~P.}\ \bibnamefont
  {Zhong}}, \bibinfo {author} {\bibfnamefont {A.}~\bibnamefont {Bienfait}},
  \bibinfo {author} {\bibfnamefont {M.-H.}\ \bibnamefont {Chou}}, \bibinfo
  {author} {\bibfnamefont {C.~R.}\ \bibnamefont {Conner}}, \bibinfo {author}
  {\bibfnamefont {{\'E}.}~\bibnamefont {Dumur}}, \bibinfo {author}
  {\bibfnamefont {J.}~\bibnamefont {Grebel}}, \bibinfo {author} {\bibfnamefont
  {G.~A.}\ \bibnamefont {Peairs}}, \bibinfo {author} {\bibfnamefont {R.~G.}\
  \bibnamefont {Povey}}, \ and\ \bibinfo {author} {\bibfnamefont {A.~N.}\
  \bibnamefont {Cleland}},\ }\href {\doibase 10.1063/5.0028840} {\bibfield
  {journal} {\bibinfo  {journal} {Applied Physics Letters}\ }\textbf {\bibinfo
  {volume} {117}},\ \bibinfo {pages} {244001} (\bibinfo {year} {2020})},\
  \Eprint {http://arxiv.org/abs/https://doi.org/10.1063/5.0028840}
  {https://doi.org/10.1063/5.0028840} \BibitemShut {NoStop}%
\end{thebibliography}%

\end{document}